\documentclass[prl,nofootinbib,twocolumn,superscriptaddress]{revtex4}

\usepackage{graphicx}
\usepackage{amsmath}
\usepackage{amsfonts}
\usepackage{amssymb}
\usepackage{dsfont}
\usepackage{dcolumn}
\usepackage{bm}
\usepackage{pstricks}
\usepackage{color}

\begin{document}

\title{ Connecting  the Direct Detection of Dark Matter 
\\ with Observation of Sparticles at the  LHC}

\author{Daniel Feldman}
\affiliation{Michigan Center for Theoretical Physics,
University of Michigan, Ann Arbor, MI 48109, USA}

\author{Zuowei Liu}
\affiliation{C.N.\ Yang Institute for Theoretical Physics, 
Stony Brook University, Stony Brook, NY 11794, USA}
\affiliation{Kavli Institute for Theoretical Physics China,
Chinese Academy of Science, Beijing, 100190, China}

\author{Pran Nath}
\affiliation{Department of Physics, Northeastern University,
 Boston, MA 02115, USA}




\begin{abstract}
An analysis is given connecting event rates for the direct detection of neutralino
dark matter with the possible signatures of supersymmetry at the LHC.
It is shown that if an effect is seen in the direct detection experiments at  a level of 
 $O(10^{-44})$ cm$^2$ for the neutralino-proton cross section, then 
within    the mSUGRA model the next heavier particle above the neutralino 
is either a stau, a chargino, or a CP odd/CP even (A/H) Higgs boson.  
Further, the collider analysis shows that models with a neutralino-proton 
cross section at the level of  $(1-5)\times 10^{-44}$ cm$^2$ 
could be probed with as little as 1 fb$^{-1}$ of integrated luminosity at the 
LHC at ~{$\sqrt s=7,10$ TeV}. The most recent limit from the five tower CDMS II  
result on WIMP-nucleon cross section is discussed in this context. 
It is argued that the conclusions of the analysis  given here are more 
broadly applicable  with inclusion of non-universalities in the SUGRA models.

 \end{abstract}

\maketitle

{\it Introduction:}
Experiments ~{for the direct detection of }  cold dark matter have made very 
significant progress recently \cite{Ahmed:2008eu,xenon,Lebedenko:2008gb} 
and are now exploring the  spin independent WIMP 
(weakly interacting massive particle)-nucleon cross sections in
the  region between  $10^{-44}-10^{-43}$ cm$^2$.
Further, the most recent five tower CDMS II result \cite{newcdms} 
has reached the sensitivity of $3.8\times 10^{-44}$ cm$^2$ 
at a mass of 70 GeV. Since the LHC has now started  its  runs
it is interesting to connect  the direct detection of dark matter
with the possible observation of sparticles at the LHC. In this paper 
we give such an analysis  within the minimal supergravity grand 
unified model (mSUGRA) \cite{Chamseddine:1982jx} and its extensions. 
In supersymmetric  (SUSY) theories with R parity the lightest supersymmetric 
particle (LSP) is absolutely stable and if  neutral it could be a candidate 
for dark matter \cite{hg}. This turns out to be the case 
when the full renormalization group analysis is carried out and 
one finds that the neutralino is indeed the LSP in most of the 
parameter space of the model \cite{lsp} and thus a  candidate for dark matter.    
The satisfaction of the relic density for the LSP neutralino can occur in a variety 
of ways. It can occur via coannihilation \cite{greist}, on the hyperbolic 
branch/focus point region (HB/FP) \cite{Chan:1997bi}, in the vicinity 
of a pole (alternately known as the funnel region) \cite{greist,dark1}, 
or in the bulk region. The coannihilation region may involve charginos, 
staus, stops, gluinos as well as other sparticles depending on the model. 
The pole region is typically in the vicinity of the $Z$ boson mass 
or the neutral CP odd/CP  even (A/H)  Higgs boson mass.

Many works have analyzed the direct detection of dark matter 
\cite{goodman,sugradark,Chattopadhyay:1998wb} 
and the predicted cross section falls within a range, 
which in large measure, is observable in the direct detection 
experiments (for a review see   \cite{jungman}).
Recently, a landscape approach to the analysis of SUGRA models 
was proposed \cite{Feldman:2007zn} and detailed analyses were 
given in  several works \cite{Feldman:2008jy,Feldman:2007fq}. 
In this ~{approach, models}  that pass the WMAP relic 
density constraints, and all the other experimental constraints,  
can be classified according to their mass  hierarchies which for practical
reasons are chosen to be  the first four sparticles excluding the lightest Higgs boson. 
In general there can be as many as $O(10^4)$ sparticle mass patterns 
for the four particle hierarchies. However, under the WMAP relic density 
and other experimental constraints only 16 such patterns survive for $\mu>0$, 
where $\mu$ is the Higgs mixing parameter in the minimal supersymmetric standard model,  
and these are labeled mSP1-mSP16 (see \cite{Feldman:2007zn} for the corresponding 
sparticle hierarchies). Further, these patterns can be put into just a few broad classes 
according to their next to the lightest supersymmetric particle (NLSP). 
Thus model points where the NLSP is a chargino are Chargino Patterns [mSP1-4], 
and similarly other models are  labeled Stau Patterns [mSP5-10], Stop Patterns [mSP11-13], 
and Higgs Patterns [mSP14-16] according to their NLSPs.

The landscape approach is also very useful in the analysis of dark matter.
Thus different patterns often generate significantly
different spin independent neutralino-proton
cross sections with the Stau, Chargino and Higgs patterns ~{producing relatively large
cross sections} while other patterns, for example, the Stop Patterns give much smaller cross sections. 
~{We explain this later in the paper.}

\begin{figure}[htbp]
  \begin{center}
  \includegraphics[width=7.5cm,height=6.5cm]{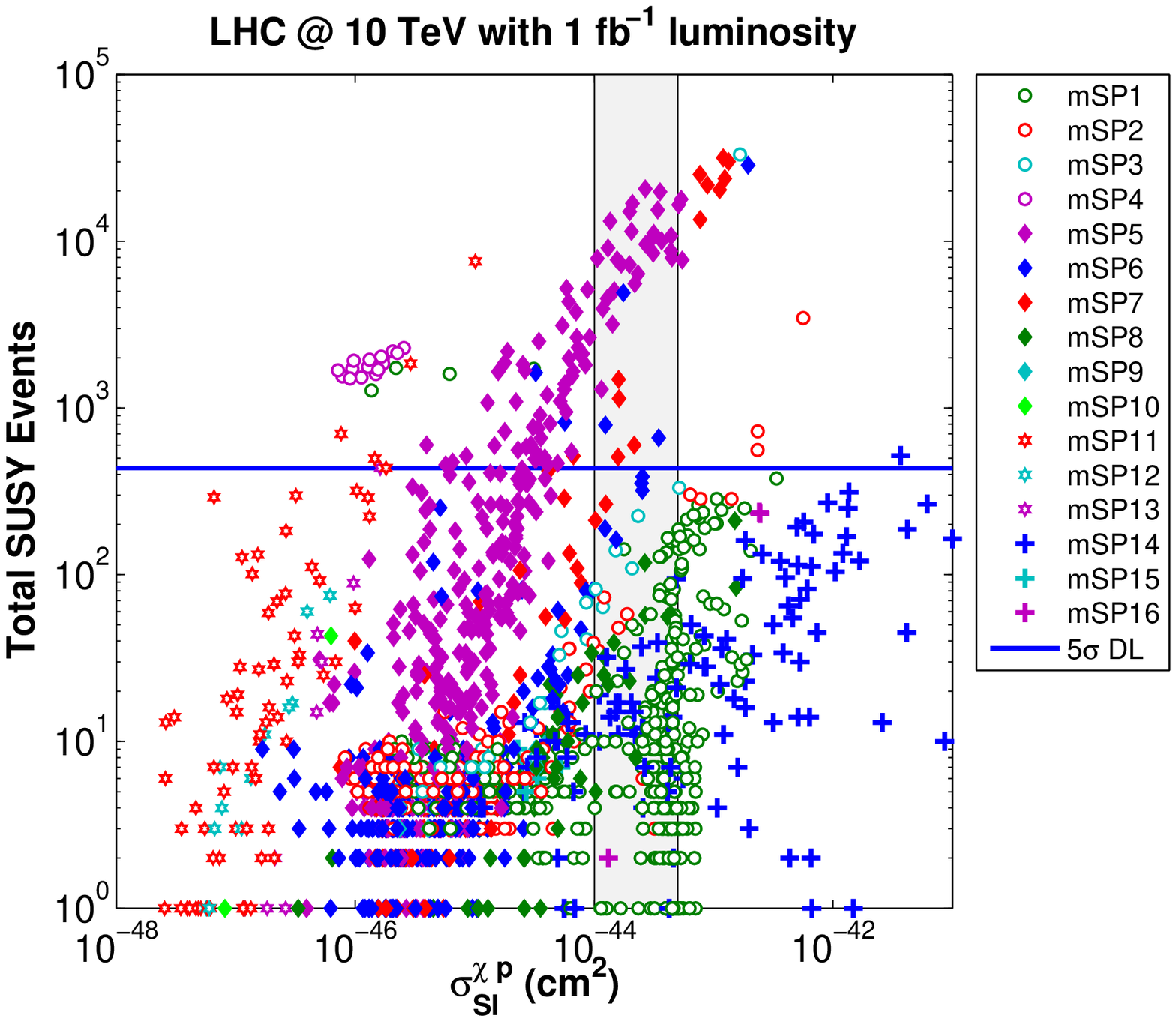}
\includegraphics[width=7.5cm,height=6.5cm]{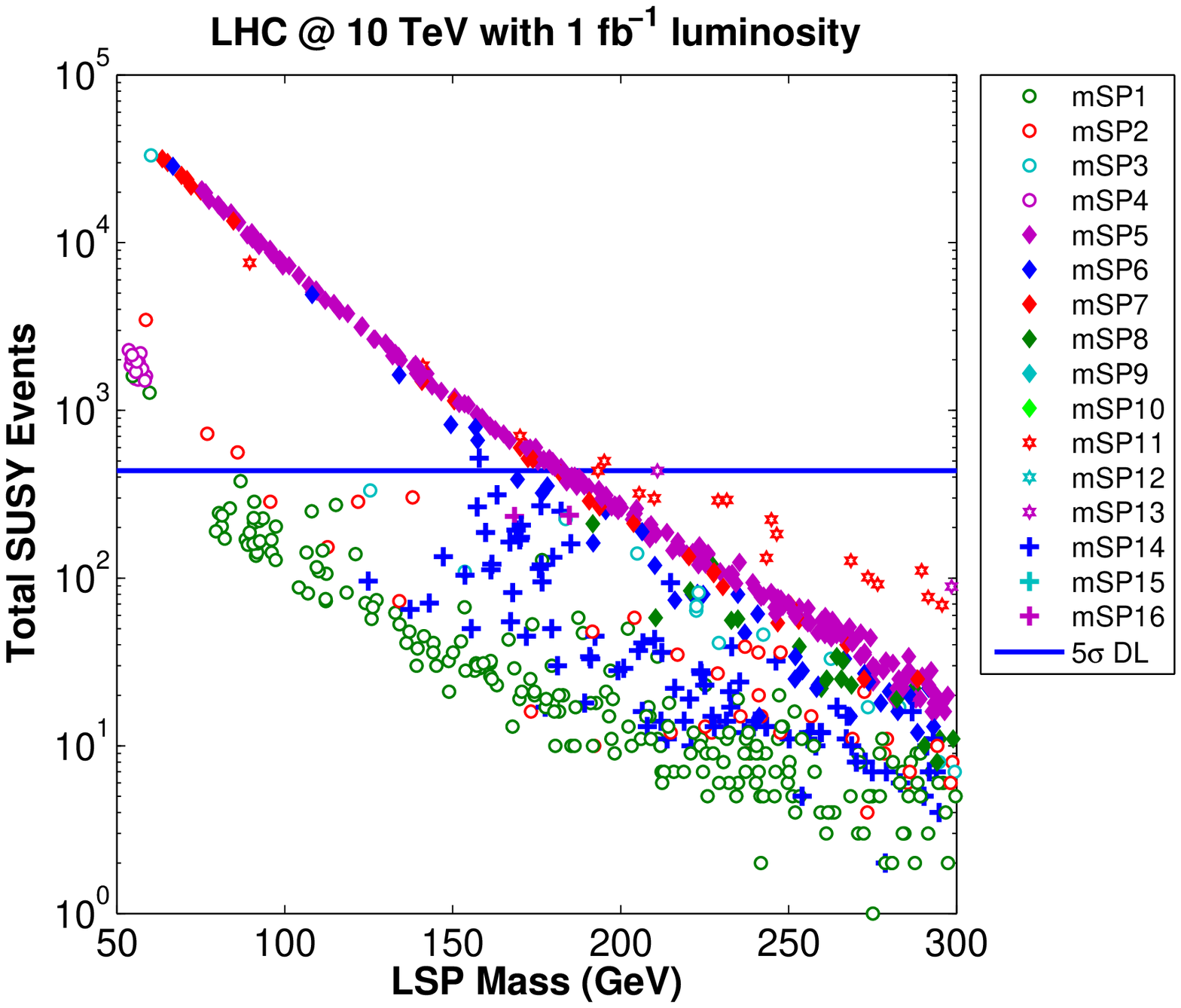}
 \includegraphics[width=7.5cm,height=6.5cm]{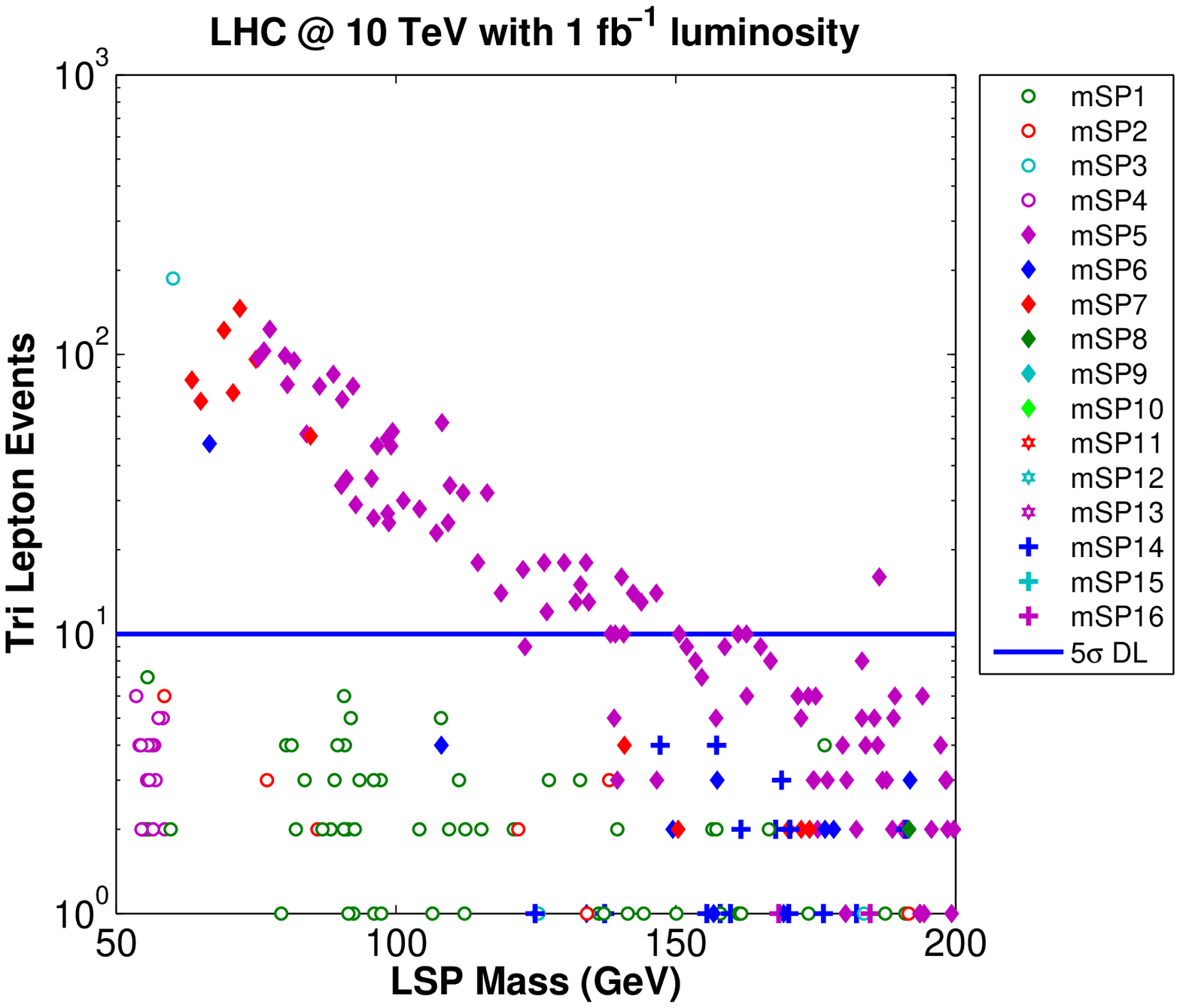}
\caption{(Color Online)   
LHC analysis at 10 TeV with 1 fb$^{-1}$ luminosity.   
Top panel: The total number of SUSY events vs the 
spin independent neutralino-proton cross section  
$\sigma^{\tilde \chi p}_{\rm SI}$.  The shaded area is the 
 region $\sigma^{\tilde \chi p}_{\rm SI}=(1-5)\times 10^{-44}$ cm$^2$.
Middle panel: The total number of SUSY events vs the LSP mass.
Bottom panel:  The number of trileptonic events \cite{trilep} vs the LSP mass.
 To suppress the background and improve the $5\sigma$ discovery reach 
of the SUSY signals (indicated by horizontal lines) we select events that have 
$\not\!\!{P_T}>200$ GeV and contain at least 2 jets with $P_T>60$ GeV. }
\label{figz}
  \end{center}
\end{figure}

{\it  Connecting the direct detection of dark matter with SUSY events at the LHC :}
LHC analyses including constraints from the relic density of cold dark matter 
have previously been investigated by many authors (see, e.g., 
\cite{Feldman:2008jy,Drees:2000he,Feng:2005gj,Arnowitt:2007nt,Kane:2007pp,Altunkaynak:2008ry} 
\cite{Baer:2008uu} \cite{Feldman:2009wv}). 
Here we investigate the relationship of  event rates in the direct detection 
experiments with the detection of sparticles at the LHC. 
Our analysis is done using  2 million candidate model
points which are then subject to the radiative electroweak symmetry 
breaking constraints, the WMAP relic density constraints, as well as 
other experimental constraints which are listed in \cite{Feldman:2007zn}. 
The parameter set  that passes these constraints is  the sample set on which 
the analysis is based. In the analysis we use micrOMEGAs \cite{micromegas}
for the relic abundance and dark matter direct detection calculations, 
and  PGS4 \cite{pgs4} for detector simulations.

\begin{figure}[htbp]
  \begin{center}
  \includegraphics[width=7.5cm,height=6.5cm]{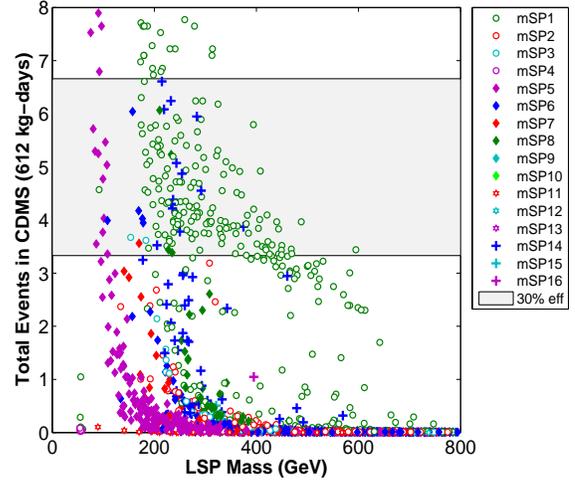}
      \caption{(Color Online) An exhibition of the number of events predicted in the CDMS detector
 with 612 kg-d of data assuming 100\% efficiency. The lower (upper) horizontal lines make the
 assumption that one (both) events in the CDMS detector are signal events and in drawing the
 lines we have assumed 30\% detector efficiency. 
}
\label{figu}
  \end{center}
\end{figure}
\begin{figure}[htbp]
  \begin{center}
   \includegraphics[width=7.5cm,height=6.5cm]{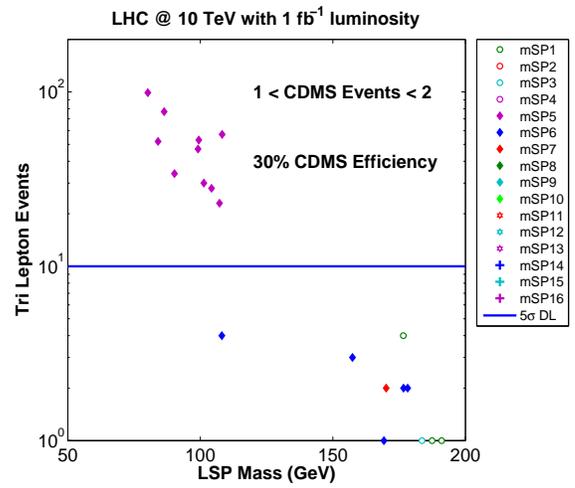} 
      \caption{(Color Online) The total number of trileptonic events at the LHC for model points 
     that lie in the corridor between the two horizontal lines in Fig.(\ref{figu}) vs the LSP mass.    
         The horizontal line indicates the $5\sigma$  discovery limit.     }
\label{figv}
  \end{center}
\end{figure}
  Fig.(\ref{figz}) gives the analysis at  $\sqrt s=10$ TeV and 1 fb$^{-1}$ of integrated luminosity.  
 To reduce the background  the following cuts were imposed: 
  $\not\!\!{P_T}>200$ GeV and  at least 2 jets with $P_T>60$ GeV. 
We discuss now the result of the analysis presented in Fig.(\ref{figz}). 
  The top panel of Fig.(\ref{figz}) gives  the total number of SUSY  events vs 
  $\sigma^{\tilde \chi p}_{\rm SI}$, while the middle panel gives  
  the total number of SUSY events vs the LSP mass, and the bottom panel gives 
 an analysis of the supersymmetric trileptonic signal \cite{trilep}. 
  The $5\sigma$ discovery limit based on the Standard Model background is exhibited in each case. 
  One finds  that a significant number of model points pass the cut and lie above the discovery limit.
     Further, most of the model points that lie in the discovery region 
     and give rise to strong WIMP-proton cross sections are those where the 
      NLSP is either a stau, a chargino, or a CP odd/CP even (A/H) Higgs boson.
\begin{figure}[htbp]
  \begin{center}
       \includegraphics[width=7.5cm,height=6.5cm]{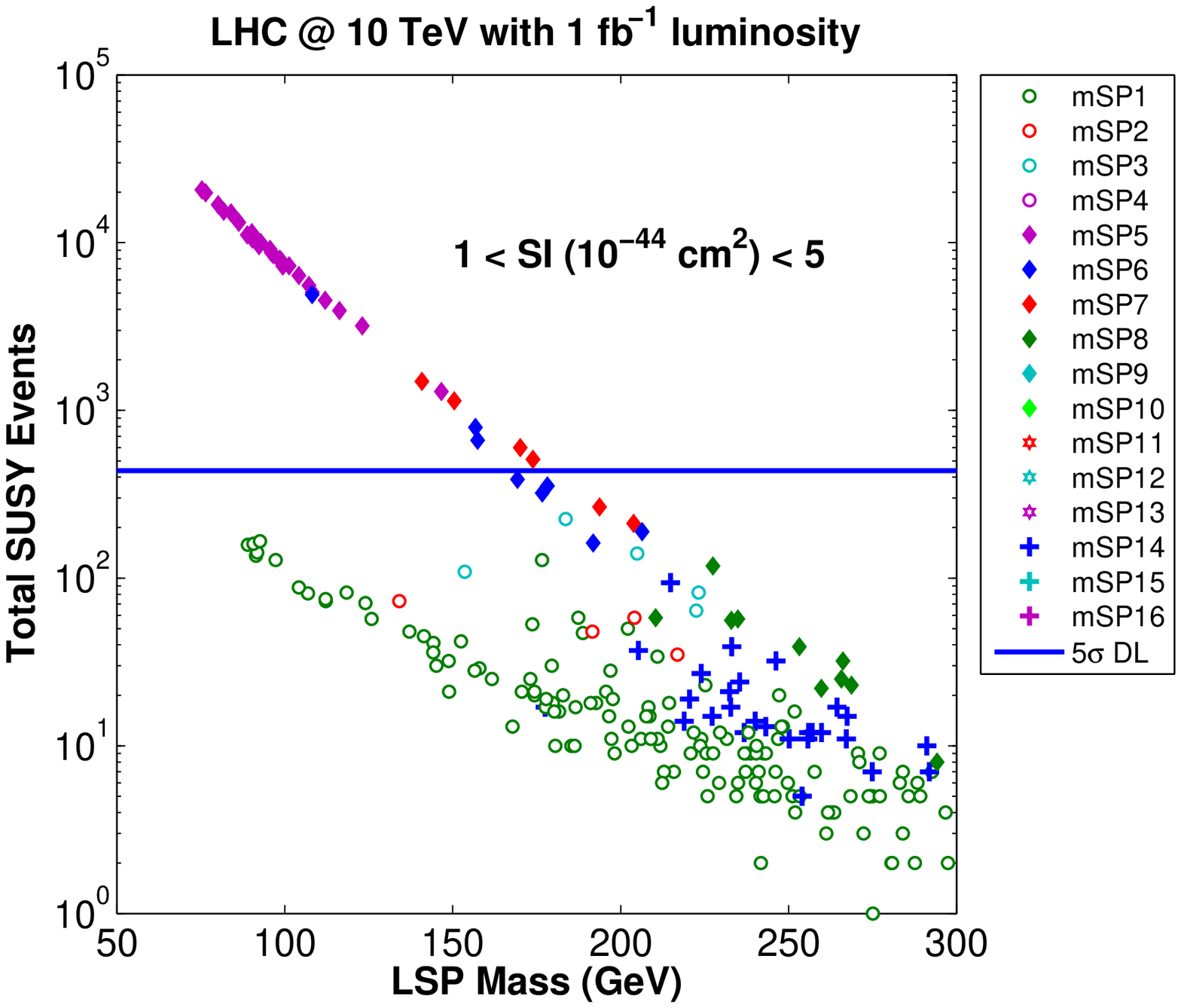}
     \includegraphics[width=7.5cm,height=6.5cm]{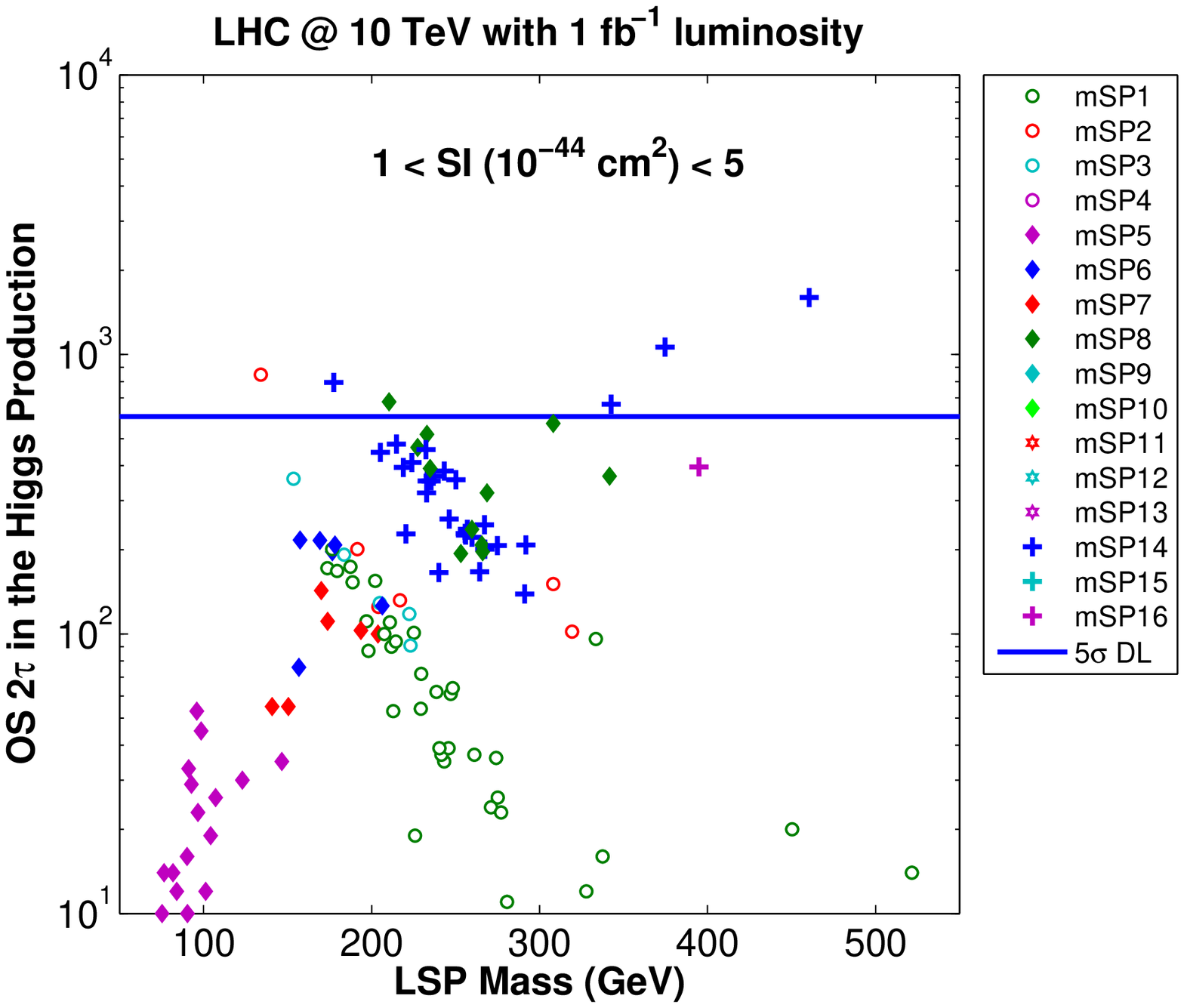}  
   \caption{(Color Online)  Top panel:  The total number of SUSY events  at the LHC
     for model points  that lie in the corridor 
     $\sigma_{\rm SI}^{\tilde  \chi p}=(1-5)\times 10^{-44}$ cm$^2$
     vs the LSP mass. The detector cuts are the same as in the analysis of Fig.(\ref{figz}).
    Bottom panel: Opposite sign (OS) two tau signal arising from Higgs production where
    only trigger level cuts are employed.
        The horizontal line indicates the $5\sigma$  discovery limit in each case.}
\label{figw}
  \end{center}
\end{figure}

We turn now to a discussion of events  in the CDMS detector for the 
models we consider in Fig.(\ref{figz}). The CDMS detector has 
accumulated 612 kg-day of data  after  quality  selection cuts are
made \cite{newcdms}. CDMS II has observed two candidate events but 
there is a possibility that both events could  be background \cite{newcdms}. 
However,  the opposite possibility that one or both of these events could be 
signal is not  excluded. It is then interesting to ask what 
the implications at the LHC would be if this were the case.
Fig.(\ref{figu}) gives the total number of  events in the CDMS detector 
vs the LSP mass for various points in the parameter space of the mSUGRA model. 
The lower (upper) horizontal lines in Fig.(\ref{figu}) correspond to one (both) CDMS II 
events being signal where we have drawn these lines assuming 30\% efficiency. 
We note that the model points  in the corridor enclosed by the two
horizontal lines have an NLSP which is either a stau, a chargino or 
an A/H Higgs boson.  This is also true in the region slightly below or 
above the corridor. Specifically we note that the patterns  
where the NLSP is a stop are not favored. 

~{Regarding the
stop NLSP, its spin independent  cross section $\sigma^{\tilde \chi p}_{\rm SI}$  is highly 
suppressed since the models that have stop NLSPs
are effectively 100 percent bino. The vanishing of the Higgsino component, therefore,
 suppresses  $\sigma^{\tilde \chi p}_{\rm SI}$.
 ~ {Removing the constraint  of   REWSB may allow one to circumvent the above feature.   However,}
  we work in the framework of REWSB, which is a key prediction of high scale models. 
There are several driving factors which generate large  $\sigma^{\tilde \chi p}_{\rm SI}$ 
 and predict large event rates at the LHC.
First, the size of the LHC signals is determined by the scale of the sparticle spectra with a lighter spectrum generally leading 
 to larger size signals. 
For the $\sigma^{\tilde \chi p}_{\rm SI}$, its largeness  is dictated 
by the relative mass scale of the mediating ~{t-channel} Higgses
and ~{s-channel} squarks, and the relative size of the LSP Higgsino component  (${\tilde H}_{i=1,2}$)
and the size of  $\tan \beta$ . For the models we study, it is the ~{t-channel}
~{Higgs exchange} and the Higgsino component of the LSP that ~{govern the spin independent cross sections}. 
Thus the models that predict the largest  $\sigma^{\tilde \chi p}_{\rm SI}$  are generally the Higgs Patterns (HPs), 
the Chargino Patterns (CPs) and the very lightest of  the  stau patterns.

\begin{figure*}[t]
  \includegraphics[width=5.75cm,height=5.5cm]{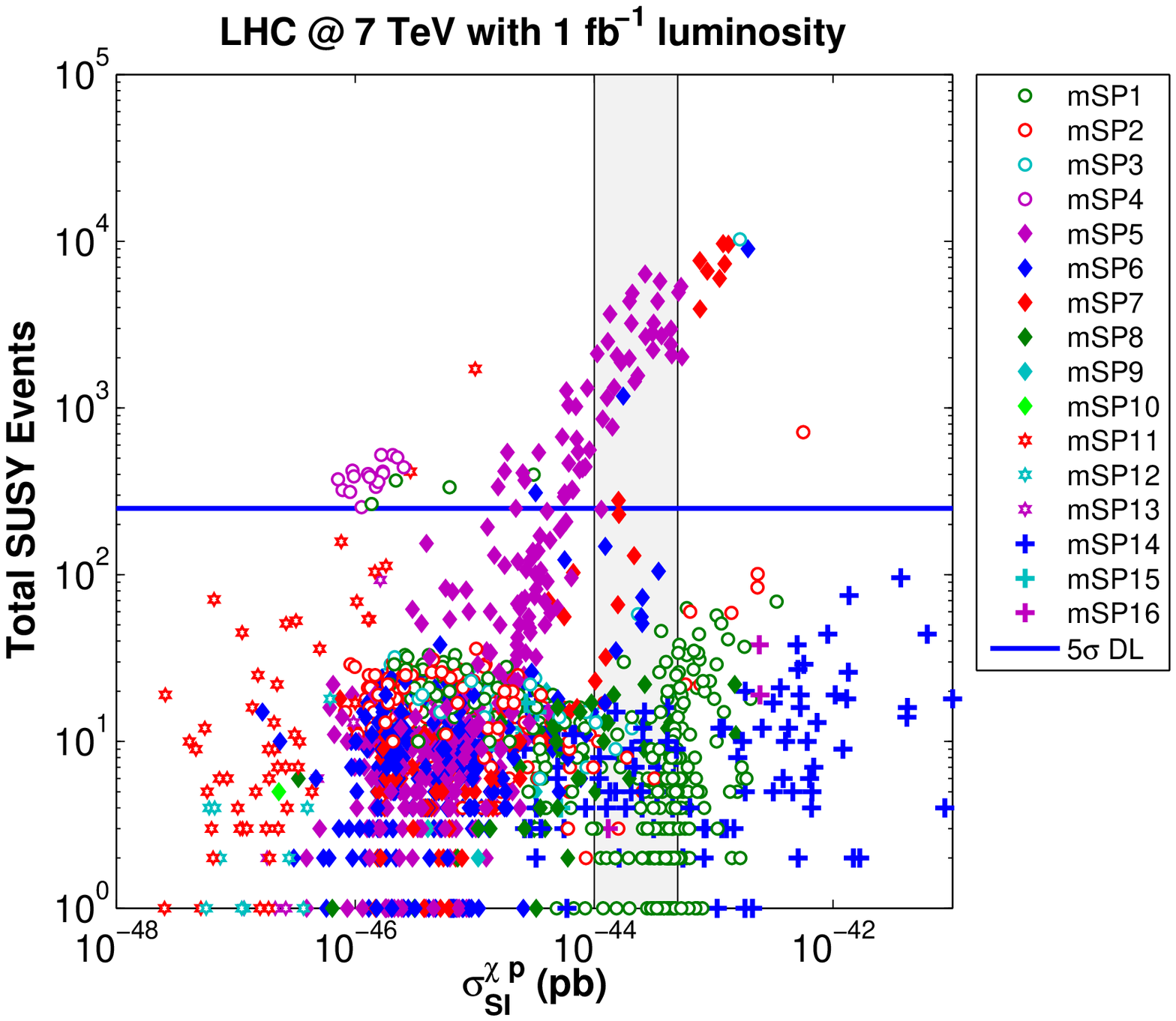}
\includegraphics[width=5.75cm,height=5.5cm]{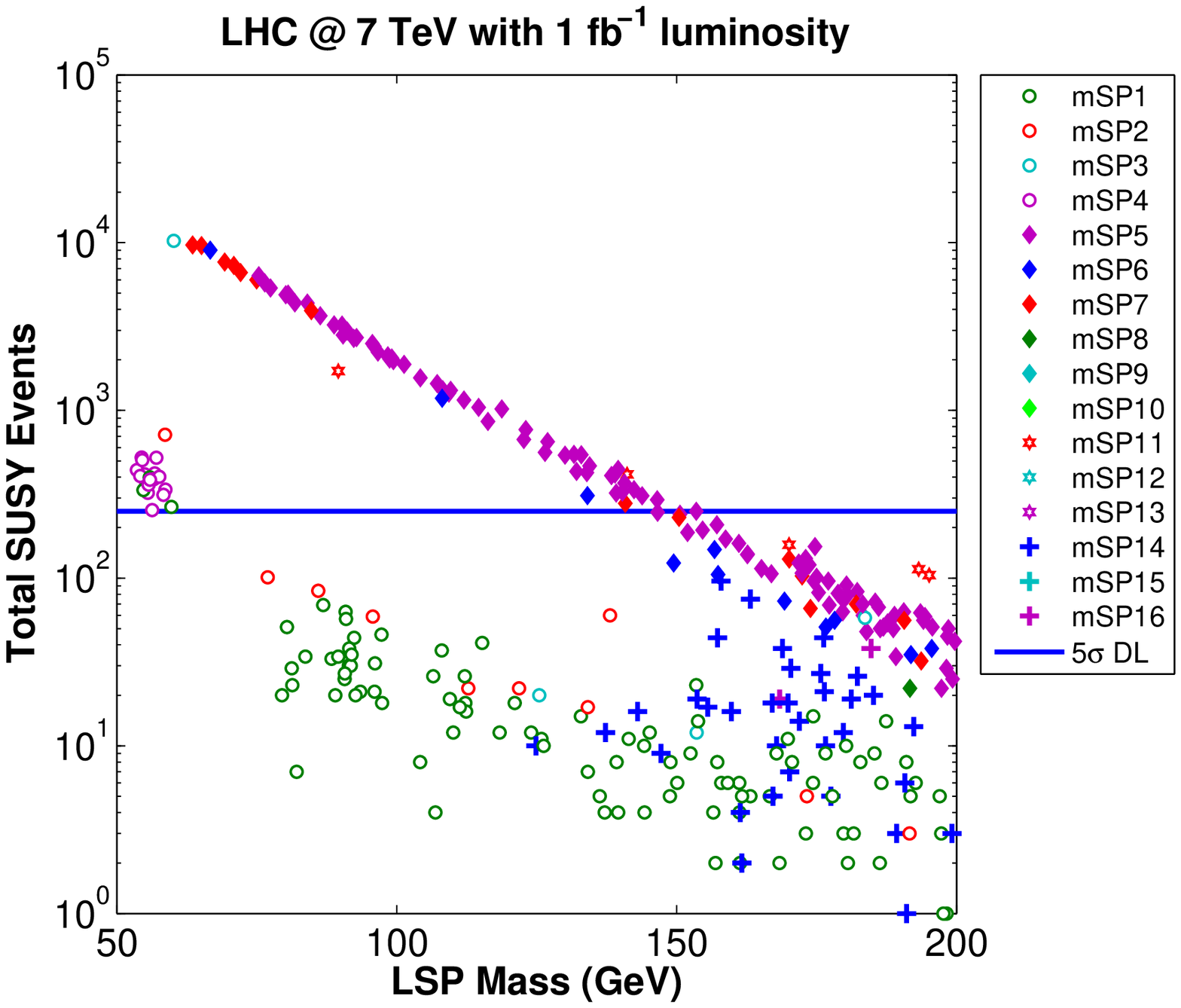}
 \includegraphics[width=5.75cm,height=5.5cm]{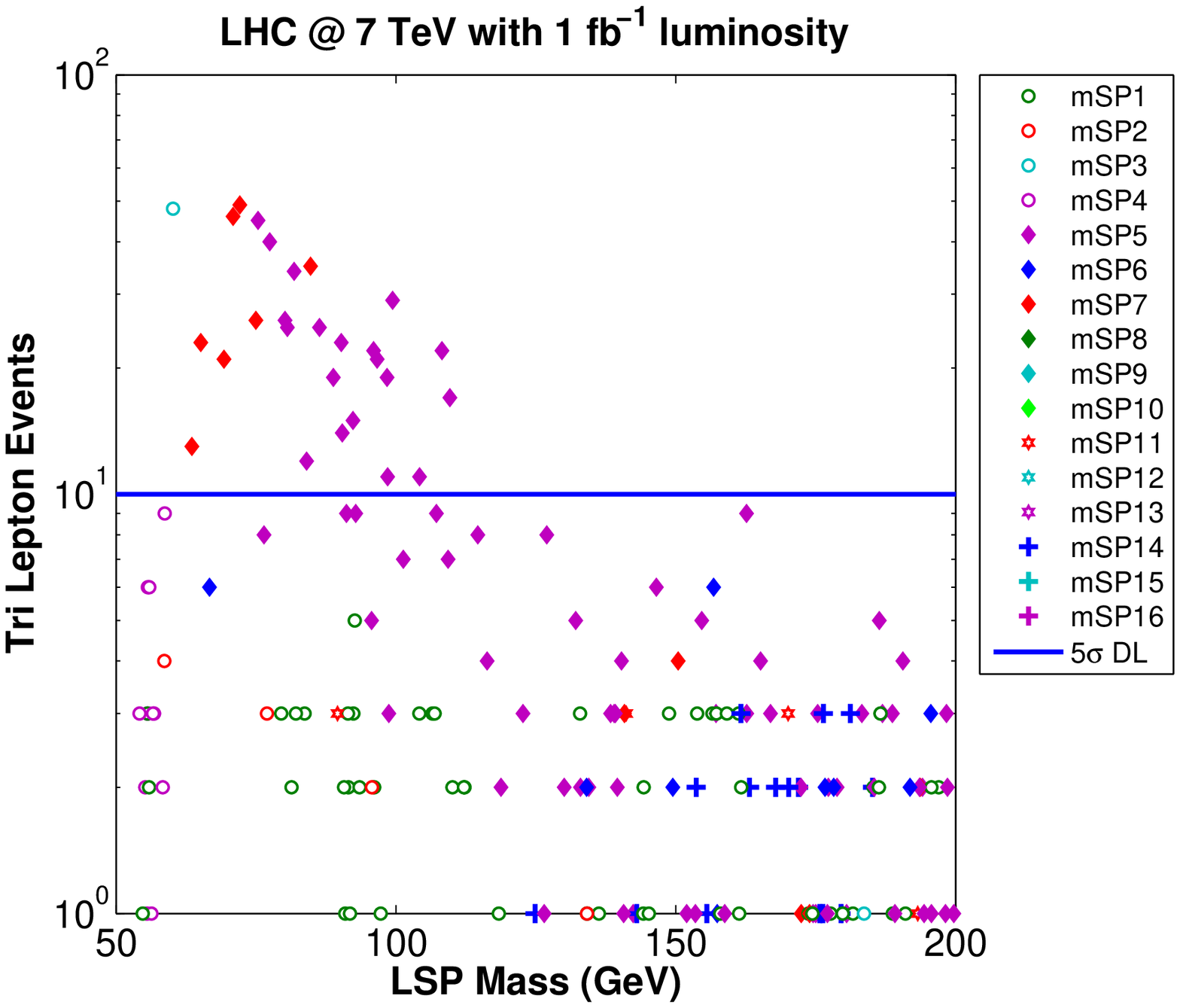}
\caption{(Color Online)   
LHC analysis at 7 TeV with 1 fb$^{-1}$ luminosity.   
Left panel: The total number of SUSY events vs the 
spin independent neutralino-proton cross section  
$\sigma^{\tilde \chi p}_{\rm SI}$.  The shaded area is the 
 region $\sigma^{\tilde \chi p}_{\rm SI}=(1-5)\times 10^{-44}$ cm$^2$.
Middle panel: The total number of SUSY events vs the LSP mass.
Right panel:  The number of trileptonic events \cite{trilep} vs the LSP mass.
 To suppress the background and improve the $5\sigma$ discovery reach 
of the SUSY signals (indicated by horizontal lines) we select events that have 
$\not\!\!{P_T}>200$ GeV and contain at least 2 jets with $P_T>60$ GeV.
These parameter points exhibited above which are discoverable at the LHC at 7 TeV via
the trileptonic signal are yet to be constrained by the Tevatron. 
\label{tev7lhc}
 }
\end{figure*}

The HPs tend to occur in the bulk of the $m_0-m_{1/2}$ plane (and are a relatively new discovery 
~{ \cite{Feldman:2007zn,Feldman:2007fq}}; ~{see \cite{Feldman:2007zn} for the large parameter space investigated
in this work and complete set of constraints}). The 
HPs have large LSP   ${\tilde H}_{i=1,2}$ components  and the  SUSY Higgses can be quite light and $\tan \beta$ is large. For the
CPs, these generally occur on the ~{Hyperbolic Branch/Focus Point region}  of REWSB. This is the region
where the scalar masses get large, but the gauginos are relatively light. Here it is the large Higgsino component and generally
a largish $\tan \beta$ which gives rise to the largeness of the  $\sigma^{\tilde \chi p}_{\rm SI}$
, while, for example, the chargino ~{LSP} coannihilation can reduce
the relic abundance to lie in the  WMAP preferred region.  Lastly, for very light stau mass patterns, it is the stau coannihilation regions
that give rise to WMAP preferred region, and for models where the LSP is light, these models can have a 
non-negligible ${\tilde H}_{i=1,2}$
components, as well as relatively ~{light CP-Even Higgs,} both of which lead to an enhanced  $\sigma^{\tilde \chi p}_{\rm SI}$.

Now, these model classes  reside in different mass hierarchies ~{\cite{Feldman:2007zn}}. It is the mass hierarchies which largely dictate
the type of signal that is visible at the LHC   (provided the mass scale is low enough to produce visible signals)
through kinematically allowed decay chains which are different for the different mass  hierarchies.
The stop patterns produce lots of jets, but are rather void of leptons. They are also ~{undetectable} with present direct detection
experiments as discussed above.  The HPs, CPs and staus, can also produce large lepton, jet and missing energy signals,
the largest missing energy coming from the HPs and the stau patterns over the CPs (see Refs. \cite{Feldman:2007zn,Feldman:2008jy}).
~{Thus in summary,}  if the model has a low SUSY mass scale of spectra it will produce a discoverable signal at the LHC.
If such a light model has enhanced  ${\tilde H}_{i=1,2}$ and/or light scalars,  it will produce a large number of events rates in
the CDMS and Xenon detectors \cite{Aprile:2010um}. The type of LHC signal is governed by the sparticle mass hierarchy which dictates the
decay chains allowed and thus the multiplicity of final  observable states. 
The above conclusions also holds more broadly in non-universal SUGRA models.}

Fig.(\ref{figv}) gives the number of trileptonic events 
vs the LSP mass for the model points that lie in the corridor enclosed by the two 
horizontal lines in Fig.(\ref{figu}).  The signature analysis 
is done at $\sqrt s=10$ at 1 fb$^{-1}$ with detector cuts as in Fig.(\ref{figz}). 
One finds that with 1 fb$^{-1}$ of integrated luminosity there is a 
significant number of model points that lie above the discovery limit 
and will be probed by the LHC. We also note that in Fig.(\ref{figv}) 
all the models that lie above the discovery limit are those with stau as the NLSP 
with mass splittings between the stau and  the LSP being $O(10\%)$ of the LSP mass. 
The relic density of these models would have arisen from 
neutralino-stau coannihilations in the early universe.
With larger luminosity other models  where the chargino 
or possibly the CP odd/CP even  Higgs is the NLSP will begin to be probed.

It is expected that by the summer of 2010 the CDMS will have 3 times 
more Germanium in their detector and thus in the near future we may 
reach a sensitivity in $\sigma_{\rm SI}^{\tilde  \chi p}$ 
of  $\sim 1 \times 10^{-44}$ cm$^2$. Anticipating this reach 
we display in Fig.(\ref{figz}) (see the shaded region in the top panel) and 
in Fig.(\ref{figw}) the SUSY signatures at the LHC in the region of the 
parameter space with $\sigma_{\rm SI}^{\tilde  \chi p}$
in the range $(1-5)\times 10^{-44}$ cm$^2$.   
Specifically,  the top panel of Fig.(\ref{figw}) shows that a significant part 
of the parameter space in this region can be probed with the total number 
of SUSY events above the $5\sigma$ discovery limit  at the LHC 
with as little as 1 fb$^{-1}$ of integrated luminosity.  
The bottom panel gives a parallel analysis for a signal from OS $2\tau$ 
events arising in the Higgs boson production modes. 
We note that  the models probed using the OS $2\tau$ from 
Higgs productions are different from the models being probed 
in the SUSY events. One also has the inverse possibility,  in that 
if the LHC sees a signal in this region, one can extrapolate 
to estimate $\sigma_{\rm SI}^{\tilde  \chi p}$ 
using the top panel of Fig.(\ref{figz}).
      
 ~{
{\it LHC Signatures at 7 TeV:} 
It has recently been announced that the LHC  will begin its early physics analysis with 
a center of mass energy  of 7 TeV and will likely operate at this energy 
till the end of 2011 or till it has accumulated 1 fb$^{-1}$  of data. 
In light of this possibility  we extend
the analysis to include a center of mass energy  of 7 TeV.  The analysis is shown in ~{Fig.(\ref{tev7lhc})}.
The results of the analysis mimic the 10 and 14 TeV results with less events
and  lower discovery limits.
Finally, one may wish to ask what more the LHC at 1 fb$^{-1}$ of integrated luminosity
can discover that the Tevatron would not. First since the LHC energy in the first run at 7 TeV 
is significantly larger than the Tevatron center of mass energy of 1.96 TeV, the
LHC would have a larger mass reach
for sparticles simply on the basis of kinematics. Further, even for model points which may also be
accessible at  the Tevatron, the LHC gives significantly larger cross sections for sparticle productions.
Thus, for example, for the mSUGRA parameter point
$m_0/{\rm GeV}= 60,  m_{1/2}/{\rm GeV}=  250,  A_0/{\rm GeV}=  -50, \tan\beta= 10$, with sign($\mu$) positive,  
which sits in mSP5, a branch of the stau coannihilation region,
one finds that for the LHC the SUSY production cross section is $\sigma_{\rm SUSY}(\sqrt{s} = 7 ~{\rm TeV}) \sim 5000 ~{\rm fb}$
while for the Tevatron, $\sigma_{\rm SUSY}(\sqrt{s} = 1.96 ~{\rm TeV}) \sim  160 ~{\rm fb}$
which indicates that the cross section at the LHC is 30 times larger than at the Tevatron for this parameter point. 
Thus  even taking account of the present larger luminosity of the
Tevatron (although the timeline for the full analysis of the data at the Tevatron relative to the LHC
early runs remains unclear), and particular cuts,  the LHC should produce a larger number of events 
in this case. 
At the same time it produces a spin independent cross section of $\sim 10^{-44} ~\rm cm^2$ with an LSP mass of 95 {\rm GeV}.
 Similar conclusions can be drawn for other SUGRA mass patterns. 
}

{\it Conclusion:} 
In this work we have given an analysis which connects the direct detection 
of dark matter with potential signals of supersymmetry at the  LHC.
Recently the sensitivities of experiments for the detection of dark matter 
have increased significantly and the current experiments are probing 
spin independent WIMP-nucleon cross sections well below the level 
of $10^{-43}$ cm$^{2}$. Thus the most recent five tower result 
from CDMS II gives an upper bound on the spin independent 
WIMP-nucleon cross section of  $3.8\times 10^{-44}$ cm$^2$  
at a WIMP mass of 70 GeV. Further, a new generation of dark matter 
experiments will in the near future begin to probe these cross sections 
at the level of $10^{-44}$ cm$^2$. It is then interesting to ask 
within the framework of supersymmetry, with neutralino as the LSP, 
what the correspondence is of a possible observation of events in a 
dark matter detector and the possible observation of SUSY signatures 
at the LHC. First an analysis of the sparticle landscape for models 
that lead to spin independent neutralino-nucleon cross section 
in the range $10^{-44}-10^{-43}$ cm$^2$ shows that the parameter 
space of mSUGRA in this range produces the NLSP which is either 
a stau, a chargino or a CP odd/CP even Higgs. Further, we have 
carried out an analysis of distinct LHC signatures for this part 
of the parameter space in the early runs. Our analysis shows 
that a part of the parameter space which gives rise to spin-independent 
neutralino-proton cross section in the range $(1-5)\times 10^{-44}$ cm$^2$ 
can produce observable signals at the LHC with as little 
as 1 fb$^{-1}$ of data at ~{$\sqrt s =7,10$ TeV}.
\\

\noindent\\\\\\
{\em Acknowledgments}:  
ZL thanks  the Kavli Institute for Theoretical Physics China (KITPC) ~{for their hospitality while completing this work.}
This research is  supported in part by DOE grant  DE-FG02-95ER40899 (MCTP),  and NSF grants PHY-0653342 (Stony Brook), and  PHY-0757959 (Northeastern University). 


\end{document}